\documentclass[%
 reprint,
superscriptaddress,
 amsmath,amssymb,
 aps,
]{revtex4-2}

\usepackage{hyperref}% add hypertext capabilities
\hypersetup{
    colorlinks=true,
    linkcolor=blue,
    filecolor=magenta,
    urlcolor=blue,
    citecolor=blue,
    bookmarks=true,
    pdfpagemode=FullScreen,
}

\usepackage{graphicx}% Include figure files
\usepackage{dcolumn}% Align table columns on decimal point
\usepackage{bm}% bold math
\usepackage{xcolor}
\usepackage{braket}
\usepackage{hyperref}% add hypertext capabilities

\begin{document}

\title{Emergence of an antiferromagnetic topological Anderson insulator in the interacting Haldane model}% Force line breaks with \\

\author{Alejandro J. Uría-Alvarez}
\email{uria@itp.uni-frankfurt.de}
\affiliation{Institut für Theoretische Physik, Goethe-Universität, 60438 Frankfurt am Main, Germany}
\author{Roser Valentí}%
\email{valenti@itp.uni-frankfurt.de}
\affiliation{Institut für Theoretische Physik, Goethe-Universität, 60438 Frankfurt am Main, Germany}
\date{\today}% It is always \today, today,

\begin{abstract}
We examine the emergence of topological Anderson insulating phases in the spinful Haldane model with Hubbard and next-neighbor density-density interactions, subject to Anderson disorder. Using finite-size exact diagonalization, we characterize the phases that arise from the interplay between topology, interactions, and disorder. In addition to standard $C=2$ topological Anderson phases, we observe an antiferromagnetic $C=1$ topological Anderson phase, consistent with the antiferromagnetic quantum anomalous Hall insulator previously identified in the clean model at finite staggered mass. We further analyze these phases using a neural network trained on the exact diagonalization data. Our results support the hypothesis that an explicit charge imbalance is required to induce the $C=1$ phase, generated by Anderson disorder rather than by a staggered mass.
\end{abstract}

\maketitle

\section{Introduction}

The joint phenomenology of topology, interactions, and disorder is key to understanding the behaviour of real materials. However, the complexity introduced by treating all three ingredients simultaneously often requires addressing them individually or pairwise. For instance, since the proposal of the first model of a topological material, the Haldane model \cite{haldane1988model}, a vast theoretical and experimental effort has been devoted to classify arbitrary clean, non-interacting materials \cite{altland1997nonstandard, Ryu_2010, alexandradinata2014wilson, Vergniory2019}. Beyond the theoretical interest of disordered topological phases,  which require dedicated tools \cite{bianco2011mapping, loring2011disordered, cerjan2022local, uria2022deep, munoz2023structural, Favata_2023, goda2006inverse, zuo2024topological}, these systems also exhibit novel phenomena, most notably disorder-induced topological phases. Such phases demonstrate that topology can persist in the presence of disorder and, in certain cases, may even arise as a consequence of disorder~\cite{Corbae_2023, agarwala2019topological, yang2019topological, marsal2020topological, wang2022, li2009topological, groth2009theory, uria2025amorphization}. Likewise, in clean, interacting topological materials, the existing machinery for the classification of non-interacting systems is no longer applicable, and must be replaced by different techniques \cite{fidkowski2010effects, fidkowski2011topological, chiu2016classification, shiozaki2018many, zhao2023failure,lessnich2021,soldini2023interacting,Herzog-Arbeitman2024, sinha2025computing, funfhaus2026manybodyeulertopology}. Similar to disorder, interactions not only renormalize single-particle properties, but can also generate interaction-driven phases that lie beyond a single-particle description, as occurs in strongly correlated systems~\cite{dagotto2005complexity, wen2017colloquium, lee2006doping}.

The existence of both interaction-induced and disorder-induced phases naturally raises the question of whether topological phases can emerge from the simultaneous action of interactions and disorder.
 
The Haldane–Hubbard model provides a minimal framework for investigating the interplay among these three aspects. Because it hosts Chern insulator (CI) phases, for which topological diagnostics are well established even in the presence of interactions and disorder~\cite{niu1985quantized,kudo2019many}, it serves as a particularly useful case study. The model, which consists of the Haldane model supplemented with Hubbard interactions and, in some cases, additional density-density interactions, has been extensively explored using a wide range of theoretical and numerical techniques~\cite{varney2010interaction, varney2011topological, he2011topological, he2012composite, vanhala2016topological, imri2016first, tupitsyn2019phase, mertz2019statistical, shao2021interplay, yi2021interplay, yuan2023phase, he2024phase, wang2024topological, shao2024phase, silva2024topological}. In the phase diagram of the clean, spinful model, i.e. the Haldane model with Hubbard interaction, it was shown that the combination of a finite staggered mass $\Delta$ and the Hubbard interaction results in an interaction-driven topological phase, the antiferromagnetic Chern insulator (AFCI), also known as antiferromagnetic quantum anomalous Hall insulator \cite{he2011topological, he2012composite}. This result was later confirmed using different methods, including Quantum Monte Carlo, exact diagonalization (ED), two-particle self-consistent method or dynamical mean-field theory \cite{vanhala2016topological, imri2016first, tupitsyn2019phase, mertz2019statistical, yuan2023phase, he2024phase}. 

The same antiferromagnetic quantum anomalous Hall insulator has also been found in other CI models, such as the Harper-Hofstadter model \cite{jiang2018antiferromagnetic, wang2019spontaneous, ebrahimkas2022antiferromagnetic, Hafez-Torbati2024, Ghasemi_2025, hafez2025explicit, wang2026antiferromagnetic}, under the same combination of staggered mass and Hubbard interaction. A nearest-neighbor repulsive density-density interaction can favor spontaneous charge order in the form of a sublattice density imbalance. Under mean-field decoupling, this spontaneous sublattice symmetry breaking produces a Semenoff-mass term, motivating the idea that the combination of both onsite Hubbard and nearest-neighbor Coulomb interactions could reproduce the same phase. However, a study of the clean extended Haldane-Hubbard model found no $C=1$ phase \cite{shao2021interplay}. The same behavior was also observed in a different model \cite{hafez2025explicit}, leading to the conjecture that explicit, rather than spontaneous, charge order is required to induce the antiferromagnetic $C=1$ phase.

On the other hand, disorder in the interacting Haldane model has been comparatively less explored. In the spinless model with repulsive density-density interactions, Anderson disorder was shown to produce a topological Anderson insulator phase \cite{yi2021interplay}, also at finite staggered mass. A particularly striking result appears when onsite Hubbard and nearest-neighbor Coulomb  interactions are combined with Anderson disorder at zero staggered mass. In Ref. \cite{silva2024topological}, mean-field theory showed that finite disorder can induce an interacting, disordered $C=1$ phase with antiferromagnetic behaviour. This would match the antiferromagnetic $C=1$ phase previously observed at finite staggered mass, but is induced by disorder rather than by the staggered mass term. Given the limitations of mean-field theory, we aim to establish the existence of this phase using ED. We then use the resulting data to train a classifier, to obtain a finer phase diagram and improve disorder statistics, which are otherwise severely limited because of the computational cost. After establishing the existence of this antiferromagnetic topological Anderson phase, we explicitly show its connection to the clean AFCI by varying both the staggered mass and the disorder strength.

\section{Model and methods}

The complete Hamiltonian for the topological interacting disordered model is given by
\begin{equation}\label{eq:mb_hamiltonian}
    H = H_0 + H_{\text{int}} + H_W,
\end{equation}
where $H_0$ is the topological single-particle model, $H_\text{int}$ contains the interaction terms and $H_W$ describes Anderson disorder. The single-particle term $H_0$ is the spinful Haldane model on the honeycomb lattice, described by the Hamiltonian \cite{haldane1988model}
\begin{align}
    H_0 &= \Delta\sum_{i,\sigma} (-1)^ic^{\dagger}_{i\sigma}c_{i\sigma} \nonumber \\ 
    &- t \sum_{\braket{i,j},\sigma}c^{\dagger}_{i\sigma}c_{j\sigma} -t_2\sum_{\braket{\braket{i,j}}, \sigma}e^{i\phi_{ij}}c^{\dagger}_{i\sigma}c_{j\sigma}.
\end{align}
The first term is the staggered potential, which takes opposite values, $\Delta(-1)^i$, on the two triangular sublattices. The hopping amplitudes $t,t_2$ correspond to first and second neighbor hopping respectively. The next-nearest neighbor hopping $t_2e^{i\phi_{ij}}$ is given in terms of the phase $\phi_{ij}=+\phi(-\phi)$ for clockwise (anticlockwise) hoppings, which fixes the chirality of the model and is responsible of the finite Chern number. In contrast with the Kane-Mele model \cite{kane2005quantum}, the spinful Haldane model consists of two identical copies of the Haldane model, and consequently displays a $C=2$ Fermi sea at half-filling in the non-interacting limit. Throughout this work, we follow previous calculations and set $t=1$, $t_2=0.2t$ \cite{vanhala2016topological, tupitsyn2019phase, shao2021interplay, yi2021interplay, silva2024topological}. We also set $\phi=\pi/2$ as it maximizes the topological region in the non-interacting topological phase diagram $(\Delta,\phi)$.

Next, we include Hubbard and nearest-neighbor density-density interactions,
\begin{equation}
    H_{\text{int}} = U\sum_{i}n_{i\uparrow}n_{i\downarrow} + V\sum_{\braket{i,j}}n_{i}n_{j}.
\end{equation}
where $n_i=n_{i\uparrow}+n_{i\downarrow}$. Finally, to evaluate the effect of disorder on the phase diagram of the system, we consider onsite Anderson disorder,
\begin{equation}
    H_W = \sum_{i,\sigma}w_ic^{\dagger}_{i\sigma}c_{i\sigma}
\end{equation}
where the onsite energies $w_i$ are sampled from a uniform distribution $w_i\in[-W/2,W/2]$, with $W$ the disorder strength.

We employ exact diagonalization to obtain the many-body ground state of Eq.~\eqref{eq:mb_hamiltonian}. Specifically, we consider the 12A cluster, which contains the $K$,$K'$ valleys at zero twist angle, following Refs.~\cite{shao2021interplay, yuan2023phase}. This cluster has $N_{\text{at}}=12$ sites, or equivalently $N_o=24$ orbitals, and is the largest cluster that can be characterized at a reasonable computational expense. The ED calculations are performed using Krylov-Shur algorithms as implemented in the \texttt{KrylovKit.jl} package \cite{Haegeman_KrylovKit_2024}.

The onsite Hubbard and the nearest-neighbour density-density interactions are responsible for inducing charge- and spin-ordered phases, namely charge-density wave (CDW) and spin-density wave (SDW) phases. We characterize both orders through staggered CDW and SDW structure factors, defined with respect to the honeycomb sublattices:
\begin{align}
    S_{\text{CDW}} &= \frac{1}{N}\sum_{i, j} (-1)^{\eta_{ij}}\braket{(n_{i\uparrow }+ n_{i\downarrow})(n_{j\uparrow} + n_{j\downarrow})}\\
    S_{\text{SDW}} &= \frac{1}{N}\sum_{i, j} (-1)^{\eta_{ij}}\braket{(n_{i\uparrow }- n_{i\downarrow})(n_{j\uparrow} - n_{j\downarrow})}
\end{align}
where $\eta_{ij}=0$ if $i,j$ belong to the same sublattice, and $\eta_{ij}=1$ otherwise. Lastly, we characterize the topological nature of the ground state by computing the many-body Chern number. This is done using twisted boundary conditions \cite{poilblanc1991twisted}, where we modify the hoppings across the periodic boundaries according to a flux insertion $\boldsymbol{\phi}=(\phi_x,\phi_y)$. The Chern number is then given by \cite{niu1985quantized}
\begin{equation} 
    C = \frac{1}{2\pi}\int d^2\boldsymbol{\phi}\Omega(\boldsymbol{\phi}),
\end{equation}
where $\Omega(\boldsymbol{\phi})$ is the Berry curvature associated with the ground state, $\Omega(\boldsymbol{\phi}) = \partial_xA_y(\boldsymbol{\phi}) - \partial_yA(\boldsymbol{\phi})$, with $\partial_{\mu}=\partial/\partial\phi_{\mu}$ and $A_{\mu} (\boldsymbol{\phi})=\braket{\Psi|\partial_{\mu}\Psi}$ the Berry connection \cite{kudo2019many}. We use the Abelian Berry connection, assuming a non-degenerate ground state. In practice, we compute the Chern number using the standard plaquette formulation of the Berry curvature, which is manifestly gauge invariant \cite{fukui2005chern, varney2011topological, Zhang_2013, kudo2019many}. The discretized Berry curvature reads
\begin{equation}\label{eq:discrete_berry_curvature}
    F_{ij}=-i\ln\left(\frac{U^x_{ij}U^y_{i+1,j}}{U^x_{i,j+1}U^y_{i,j}}\right),
\end{equation}
where
\begin{equation}
    U^x_{ij} = \frac{\braket{\Psi_{ij}|\Psi_{i+1,j}}}{\left|\braket{\Psi_{ij}|\Psi_{i+1,j}}\right|}\quad 
    U^y_{ij} = \frac{\braket{\Psi_{ij}|\Psi_{i,j+1}}}{\left|\braket{\Psi_{ij}|\Psi_{i,j+1}}\right|}.
\end{equation}
Here, $\ket{\Psi_{ij}}$ is the ground state of the Hamiltonian with flux insertions $(\phi_x,\phi_y)_{ij}=(\frac{i}{N_x},\frac{j}{N_y})2\pi$, such that $i\in[0,N_x)$ and $j\in[0,N_y)$. The Chern number is then obtained as $C=\frac{1}{2\pi}\sum_{i,j}F_{ij}$.

Although exact diagonalization provides direct access to the many-body ground state, it is computationally expensive and therefore limits the number of disorder realizations that can be averaged over. To address this limitation, we train a classifier to improve the disorder statistics and resolve the phase diagram in finer detail. The classifier is a convolutional neural network (CNN), that takes the single-particle density matrix (SPDM) as input and predicts the Chern number. The possible classes are either 0 for SDW, CDW or trivial phases; C=2 for the conventional CI or C=1  for the AFCI. The network is trained using the \texttt{PyTorch} package \cite{Ansel_PyTorch_2_Faster_2024}.

To improve the quality of the predictions, we performed a small amount of feature engineering on the input data. Instead of providing only the SPDM at zero twist angle, we evaluate it at different twists:
\begin{equation}
    \rho_{ij}(\boldsymbol{\phi})= \braket{\Psi(\boldsymbol{\phi})|c^{\dagger}_ic_j|\Psi(\boldsymbol{\phi})}
\end{equation}
This choice is motivated by the fact that the Chern number is obtained from the Berry curvature over twist-angle space. Hence, the ground state and consequently the SPDM, will carry more information about topology when evaluated at different twists. In particular, we restrict ourselves to the high symmetry points of the twist-angle Brillouin zone $[0,2\pi]^2$. From testing,we use $\{\rho(\mathbf{0}),\rho(\boldsymbol{\pi})\}$, which increases the accuracy by approximately 0.5\% over using only $\rho(\mathbf{0})$. Including additional twist angles did not meaningfully improve the network accuracy.

\section{Results}

\begin{figure}[t]
    \centering
    \includegraphics[width=1\linewidth]{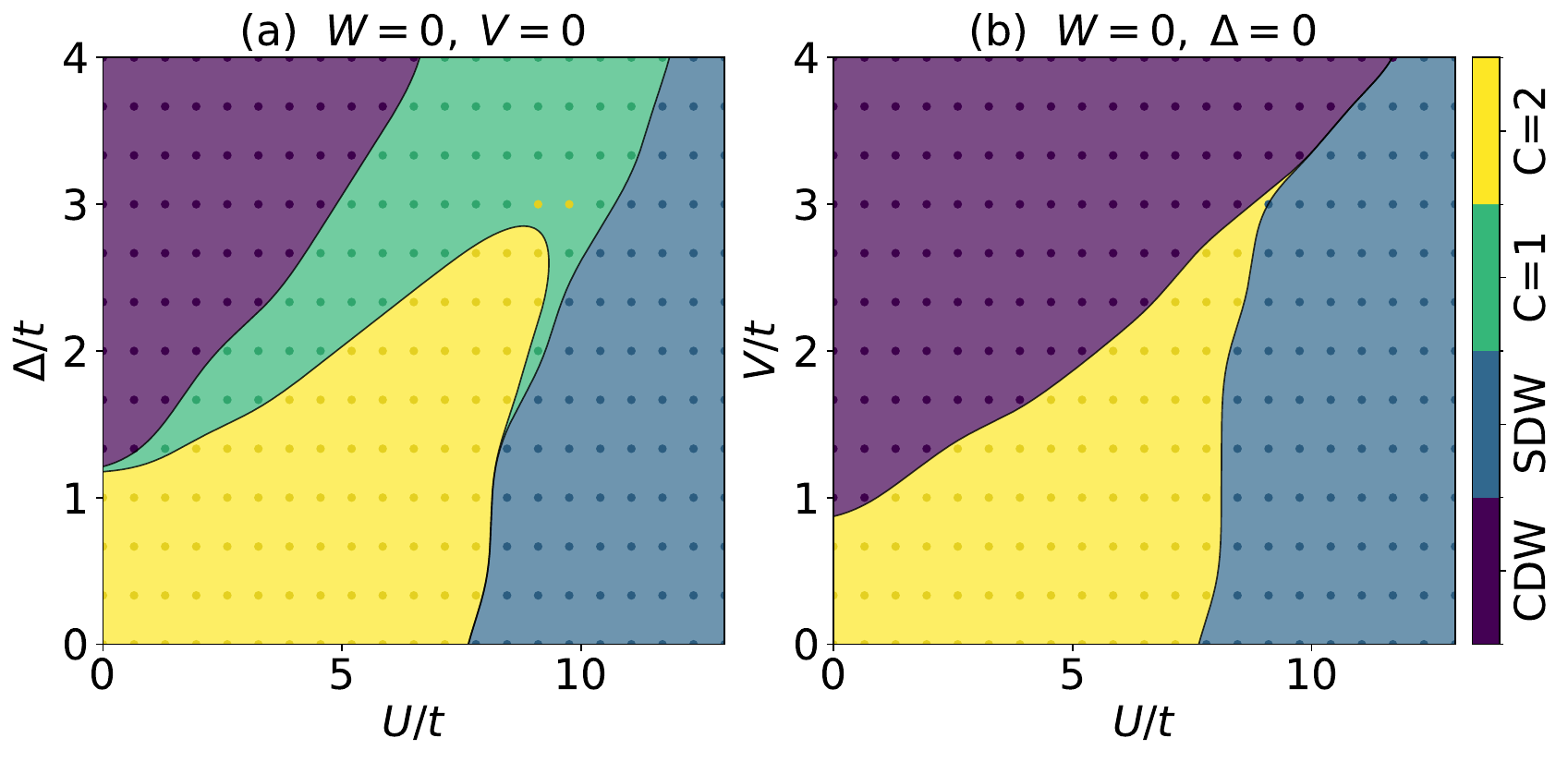}
    \caption{(a) Phase diagram of the clean extended Haldane-Hubbard model as a function of the staggered mass $\Delta$ and the Hubbard interaction strength $U$, at $V=0$. (b) Phase diagram of the clean extended model as a function of $U$, $V$, at zero staggered mass $\Delta=0$. The colored points denote the ED results and the grid used in the calculations, whereas the smooth lines marking the boundaries between phases were obtained by interpolating the points.}
    \label{fig:fig1}
\end{figure}

\begin{figure}[t]
    \centering
    \includegraphics[width=1\linewidth]{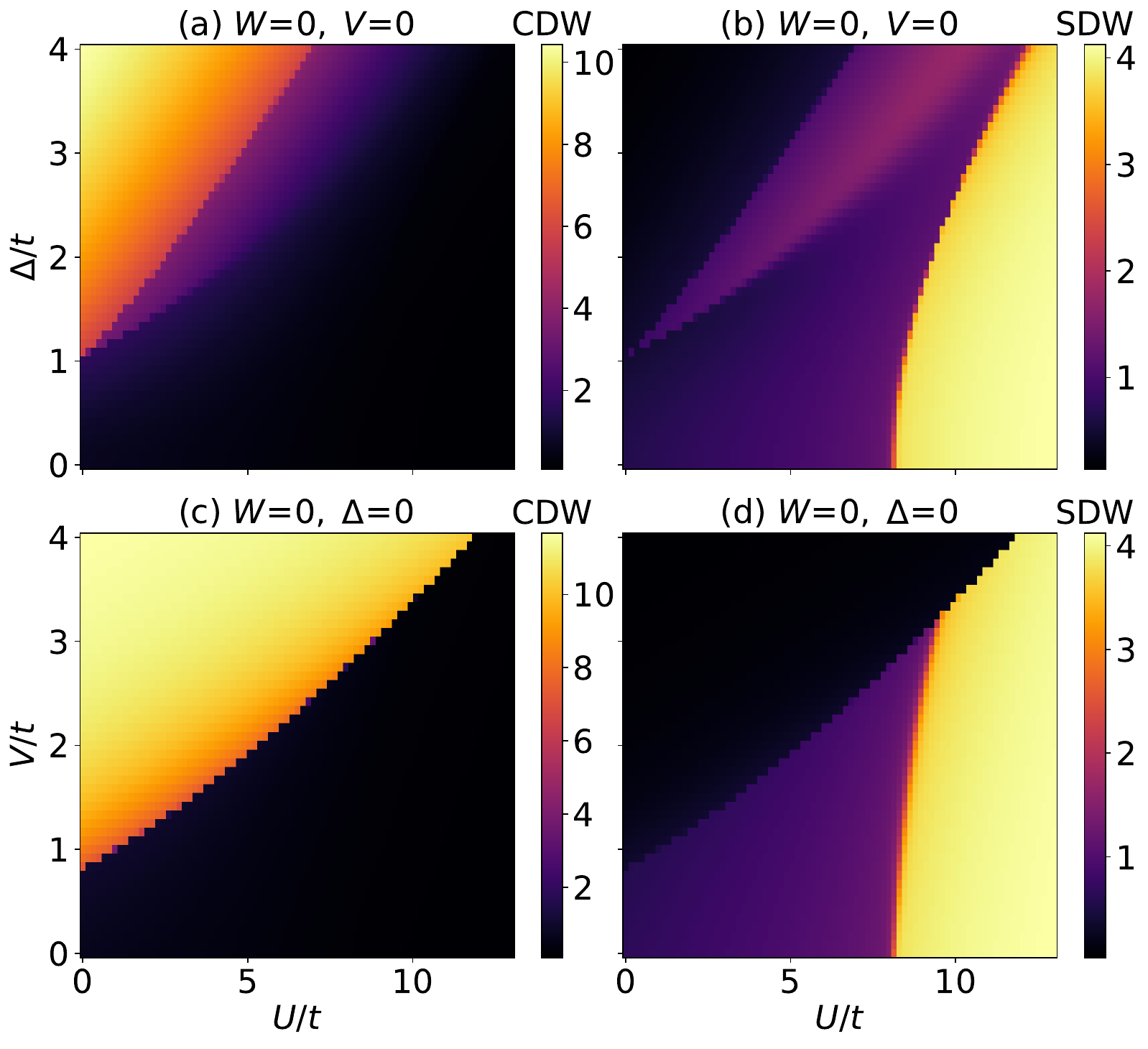}
    \caption{CDW and SDW diagrams of the clean extended Haldane-Hubbard model. (a, b) Diagrams as a function of $(\Delta, U)$ at $V=0$. (c, d) Diagrams as a function of $(U,V)$ at $\Delta = 0$.}
    \label{fig:fig2}
\end{figure}

We first assess the behavior of the Haldane-Hubbard model at zero disorder. The AFCI was originally observed at finite staggered mass $\Delta$ and finite $U$ \cite{he2011topological, he2012composite}. As discussed above, using ED for the 12A cluster, we obtain the $(\Delta, U)$ phase diagram at $V=0$, shown in Fig.~\ref{fig:fig1}(a). This diagram shows the existence of a CDW phase induced by the staggered mass, as it favors occupation of one sublattice over the other. At the opposite side of the phase diagram, we find a SDW phase, corresponding to the antiferromagnetic phase typically induced by the Hubbard interaction \cite{Sorella2012, rachel2010topological, lessnich2024spin}. The yellow region corresponds to the interacting CI with $C=2$, which can be understood as two identical spin copies of the non-interacting CI. The emergent phase, previously reported in several works, is the green $C=1$ region. This phase has been identified as an AFCI, as it exhibits a finite SDW component while having $C=1$. This can be seen explicitly in Fig.~\ref{fig:fig2}(a, b), where we show the CDW and SDW components of the $(U,\Delta)$ diagram. In this case, the CDW and SDW can be also identified from the Chern number, since they correspond to well-separated $C=0$ regions. The $C=1$ phase exhibits both small but finite CDW and SDW character. Since quantum anomalous Hall states are typically realized in ferromagnetic materials \cite{rui2010quantized, yujun2020quantum, serlin2020intrinsic, Liu2023}, we emphasize the finite SDW component and label the present phase as an AFCI.

Since the staggered mass term induces both a CDW phase and the antiferromagnetic topological phase, one might expect a first-neighbour density-density interaction to produce similar physics, as electrons would tend to occupy one of the two sublattices to reduce the total energy. This possibility was originally assesed in Ref.~\cite{shao2021interplay}. Here, we reproduce the same phase diagram as a function of $U$ and $V$ at zero staggered mass, $\Delta=0$, shown in Fig.~\ref{fig:fig1}(b). While the CDW, SDW and $C=2$ phases appear following the same reasoning as before, the $C=1$ phase is absent. This behaviour was also found in the Harper-Hofstadter model in Ref.~\cite{hafez2025explicit}, leading to the proposal that the $C=1$ phase requires explicit charge order, as produced by the staggered mass, rather than spontaneous charge order generated by the Coulomb $V$ term. As before, the phases are characterized through the explicit computation of the order parameters, shown in Fig.~\ref{fig:fig2}(c, d), and the Chern number.

\begin{figure}[t]
    \centering
    \includegraphics[width=1\linewidth]{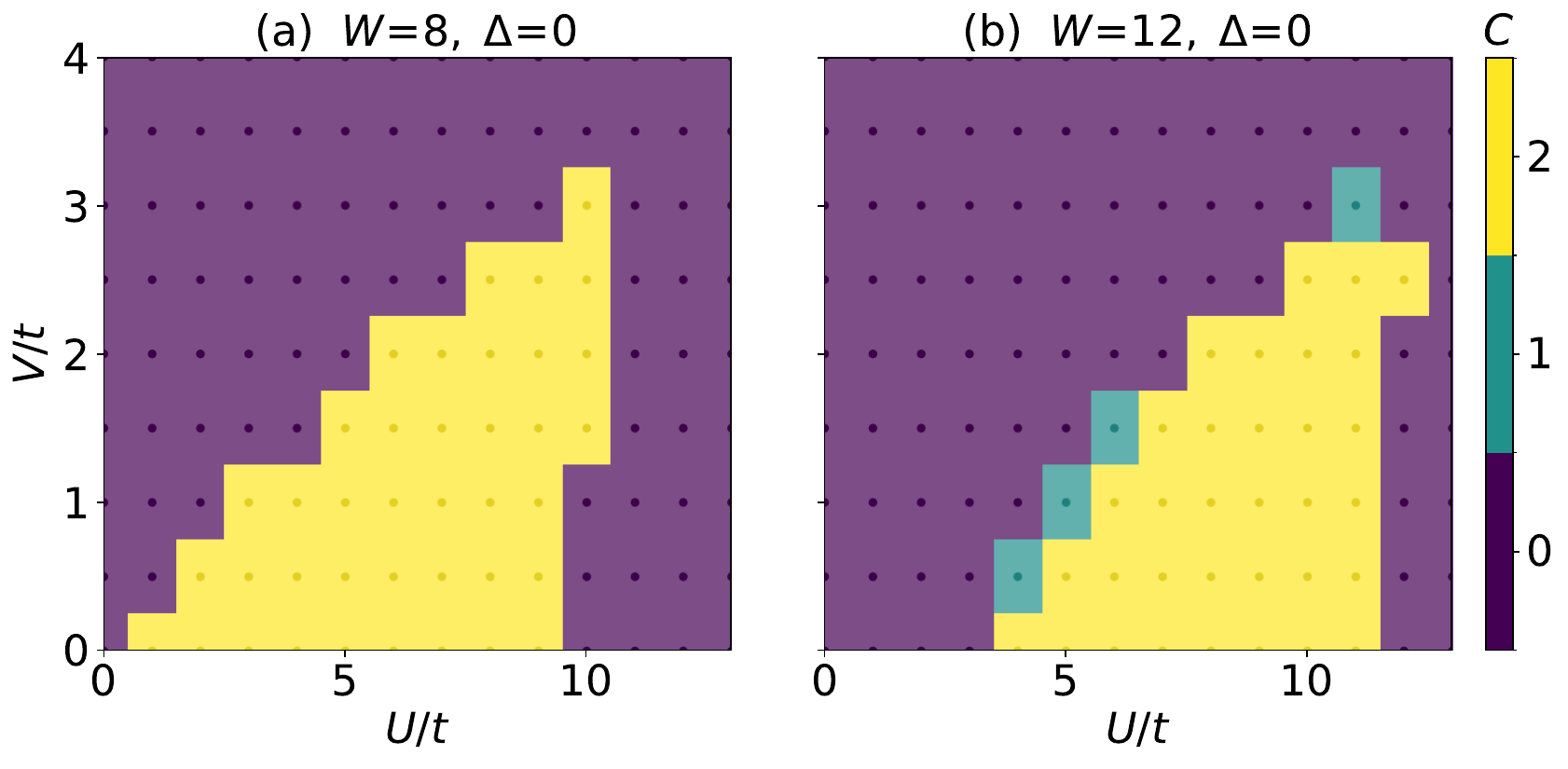}
    \caption{Phase diagram of the disordered extended Haldane-Hubbard model at disorder strenghts (a) $W=8$ and (b) $W=12$. Beyond the displacement of the $C=2$ region towards higher values of $U$, at $W=12$ we observe the emergence of an intermediate $C=1$ phase. Both diagrams are averaged over $N_s=20$ disorder realizations.}
    \label{fig:fig3}
\end{figure}

Thus, in the clean extended Haldane-Hubbard model, the antiferromagnetic quantum anomalous Hall insulator does not appear at zero staggered mass. It was later shown, however, that upon the introduction of Anderson disorder the model can develop a $C=1$ phase with antiferromagnetic character \cite{silva2024topological}. Since these results were obtained using mean-field theory, we want to confirm the existence of this phase in the presence of strong correlations using ED. Following the clean calculations, we diagonalize the 12A cluster at different disorder strenghts $W\in \{1,\ldots,12\}$ as a function of $(U,V)$ and characterize the resulting ground states using the CDW and SDW structure factors and the Chern number. We show the results for two specific disorder strengths, $W=8$ and $W=12$, in Fig.~\ref{fig:fig3}. These topological phase diagrams are averaged over $N_s=20$ disorder realizations, and we represent the majority Chern number, i.e. the modal Chern number of the distribution at each point.

There are two main obsevations: at intermediate disorder, $W=8$, Fig.~\ref{fig:fig3}(a), the $C=2$ regions shifts to higher values of $U$. This produces a trivial region around $(U,V)=(0,0)$, which can be understood as the trivial Anderson insulator. More importantly, values of $U$ that were topologically trivial at $W=0$ become topological at finite disorder, realizing a topological Anderson insulator as a disorder-induced topological phase. This behavior was already reported in Refs.~\cite{yi2021interplay, silva2024topological} for the spinless and the spinful disordered models, respectively. 

The second observation is that, at stronger disorder, $W=12$, Fig.~\ref{fig:fig3}(b), a $C=1$ region appears at the boundary between the trivial and the topological regions, in addition to the displacement of the $C=2$ region toward higher values of $U$. Besides the top pocket, already predicted in Ref. \cite{silva2024topological}, we also find an additional pocket, suggesting a longer extent of the $C=1$ region. Since the mesh used for the predictions is relatively coarse and the number of disorder realizations is limited, we use the generated data to train a classifier that predicts the Chern number from the SPDM evaluated at two twist angles. 

While one could use the full many-body ground state as input to the network, this representation is impractical for training, given that for the 12A cluster a single ground-state vector requires $\sim 2^{24}\cdot128=256$ MB of storage. Therefore, an intermediate representation that reduces the storage requirement is mandatory. We opt for the SPDM, which is known to contain information about the topology \cite{hannukainen2024interacting, zheng2020measuring, alexandradinata2011trace, turner2010entanglement} and has already been used in machine-learning methods to predict topological behaviour in interacting and non-interacting systems \cite{carvalho2018real, che2020topological, uria2022deep, simone2023unsupervised, uria2025amorphization}. 

As seen in Fig.~\ref{fig:fig3}, the $C=1$ is scarce compared with the majority classes $C=2$ and $C=0$. Since the minority $C=1$ class is the relevant one, this constitutes a class-imbalance problem, requiring additonal care during training to ensure reliable detection of this phase. We refer to the Supplemental Material for details on the network architecture and training parameters; here, we used a focal loss \cite{lin2020focal, Johnson2019} which was introduced to improve the detection of minority classes. This yields a $F_1$ score of 90\% for the $C=1$ class on the complete dataset.

\begin{figure}[t]
    \centering
    \includegraphics[width=1\linewidth]{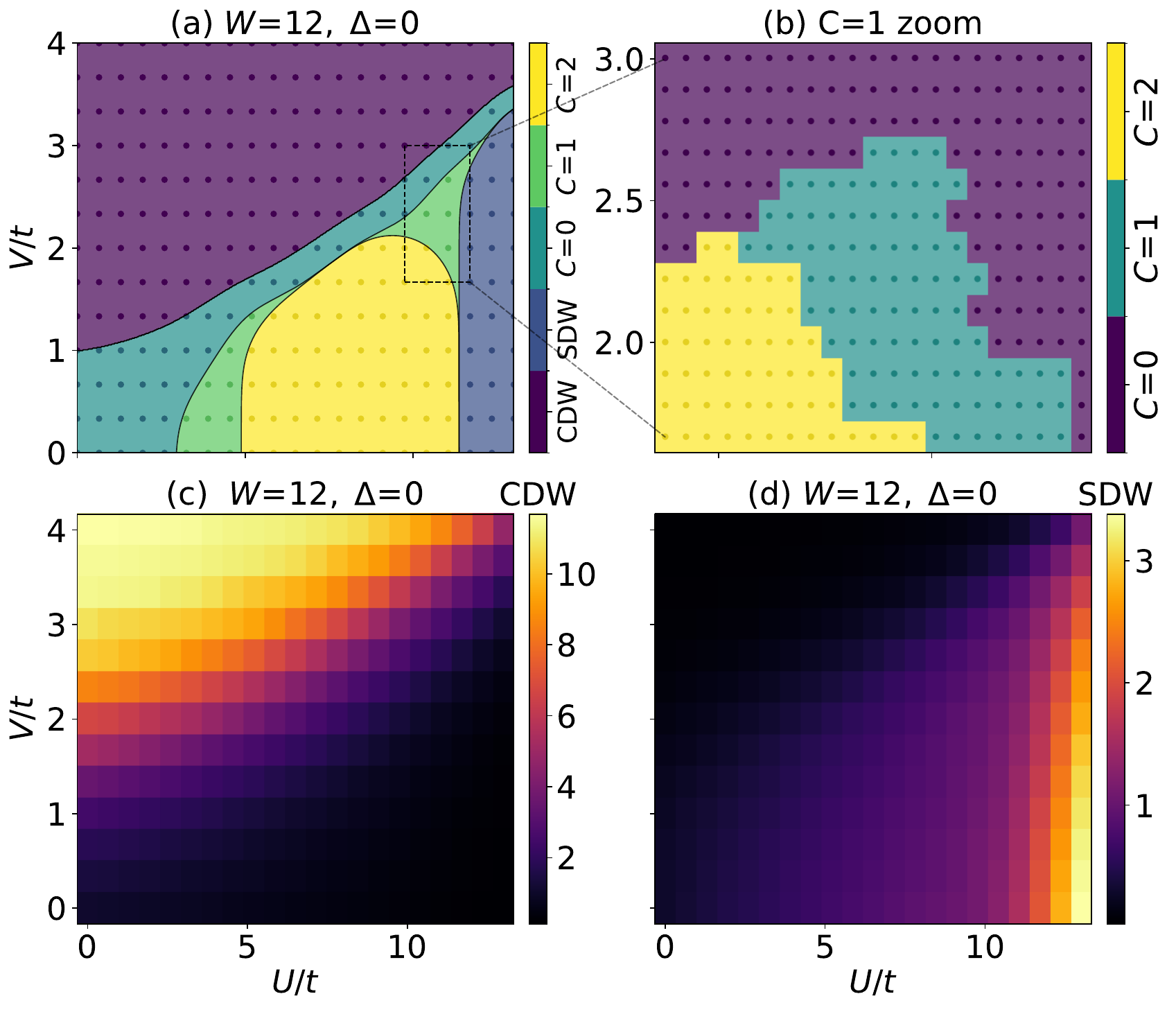}
    \caption{(a) $(U, V)$ phase diagram  of the disordered extended Haldane-
Hubbard predicted by the CNN. Beyond the conventional CDW, SDW, and $C=2$ phases, the classifier identifies two well-defined pockets for the $C=1$ phase, together with an intermediate $C=0$ phase that does not correspond to the fully developed CDW or SDW phases. (b) Zoom-in of the upper $C=1$ pocket. In both figures, the dots denote the discrete grid used to evaluate the phase diagrams, whereas the smooth boundary lines in (a) were obtained interpolating the points. (c, d) CDW and SDW diagrams as a function of $U$ and $V$. All plots have been disorder-averaged over $N_s=100$ realizations.}
    \label{fig:fig4}
\end{figure}

Using the trained network, we predict the phase diagram at $W=12$ across $N_s=100$ disorder realizations. The predictions are shown in Fig.~\ref{fig:fig4}(a), where we plot the majority phase diagram as a function of $(U,V)$. As before, this diagram is obtained using the CDW and SDW structure factors, see Fig.~\ref{fig:fig4}(c, d), together with the network-predicted Chern number. The classifier reproduces the same main regions as the ED results, namely the shifted $C=2$ region and the two pockets corresponding to the $C=1$ phase. From the SDW diagram, it is clear that both $C=1$ regions exhibit a finite SDW component and can therefore be regarded as an antiferromagnetic topological Anderson insulator, since they are induced by disorder. Following the hypothesis introduced in Ref.~\cite{hafez2025explicit}, we argue that local Anderson disorder plays a role analogous to that of the staggered mass, by generating an explicit charge imbalance that allows the emergence of the $C=1$ phase.

To verify the finite extent of the $C=1$ region, we zoom in on the upper pocket, represented in Fig.~\ref{fig:fig4}(b). There, we show the predicted majority Chern number over $N_s=100$ disorder realizations, further supporting the existence of this region. In addition to the four main phases, we introduce a fifth phase, denoted in Fig.~\ref{fig:fig4}(a) as the $C=0$ phase. It corresponds to the trivial Anderson insulator, this is, a disorder-induced trivial region that cannot be assigned a purely CDW or SDW character. As seen in Figs. \ref{fig:fig4}(c, d), it exhibits both order parameters, but neither is sufficiently large to identify the phase with the main CDW or SDW regions.

\begin{figure}[t]
    \centering
    \includegraphics[width=1\linewidth]{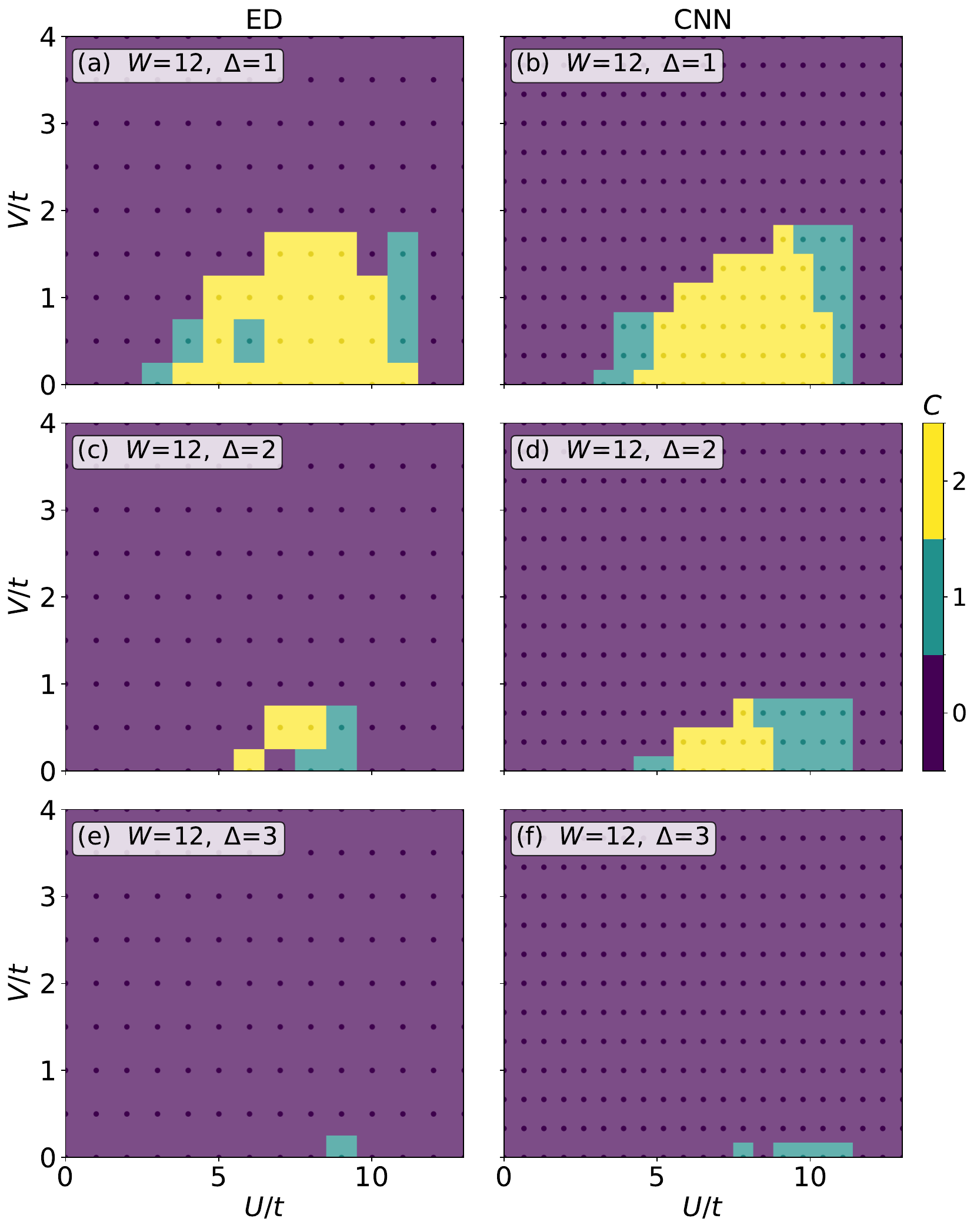}
    \caption{Comparison between (left) the ED results and (right) the convolutional neural network predictions 
    of the disordered extended Haldane-Hubbard at disorder strength $W=12$ and different staggered masses. (Top row) $\Delta=1$, (middle) $\Delta=2$ and (bottom) $\Delta=3$. The ED results have been disorder-averaged over $N_s=15$ realizations, while the CNN predictions were averaged over $N_s=100$ realizations.} 
    \label{fig:fig5}
\end{figure}

With this, we have established the existence of an antiferromagnetic topological Anderson phase at zero staggered mass. However, in the clean model, this same phase appears at finite staggered mass. Since both phases exhibit the same Chern number and SDW character, we expect them to be smoothly connected in parameter space. Starting from the phase diagram at $W=12, \Delta=0$, we compute the topological phase diagrams at $W=12$ and $\Delta\in\{1,2,3,4,5\}$ as a function of $(U,V)$. The results are shown in Fig.~\ref{fig:fig5}, where we compare the ED results in the left column, with the neural network predictions,  right column. We note that the neural network used here is the same as before; it was trained on the full ED dataset, which includes multiple values of $(U,V,W,\Delta)$. Both methods predict qualitatively the same regions, with the $C=2$ and the $C=1$ regions shrinking as the staggered mass increases. This suggests a competition between $W$ and $\Delta$: both induce explicit charge order in the system, but in asymmetric ways that do not simply add constructively. For $\Delta > 3$, both topological phases disappear and the diagram becomes entirely trivial. 

\begin{figure}[t]
    \centering
    \includegraphics[width=1\linewidth]{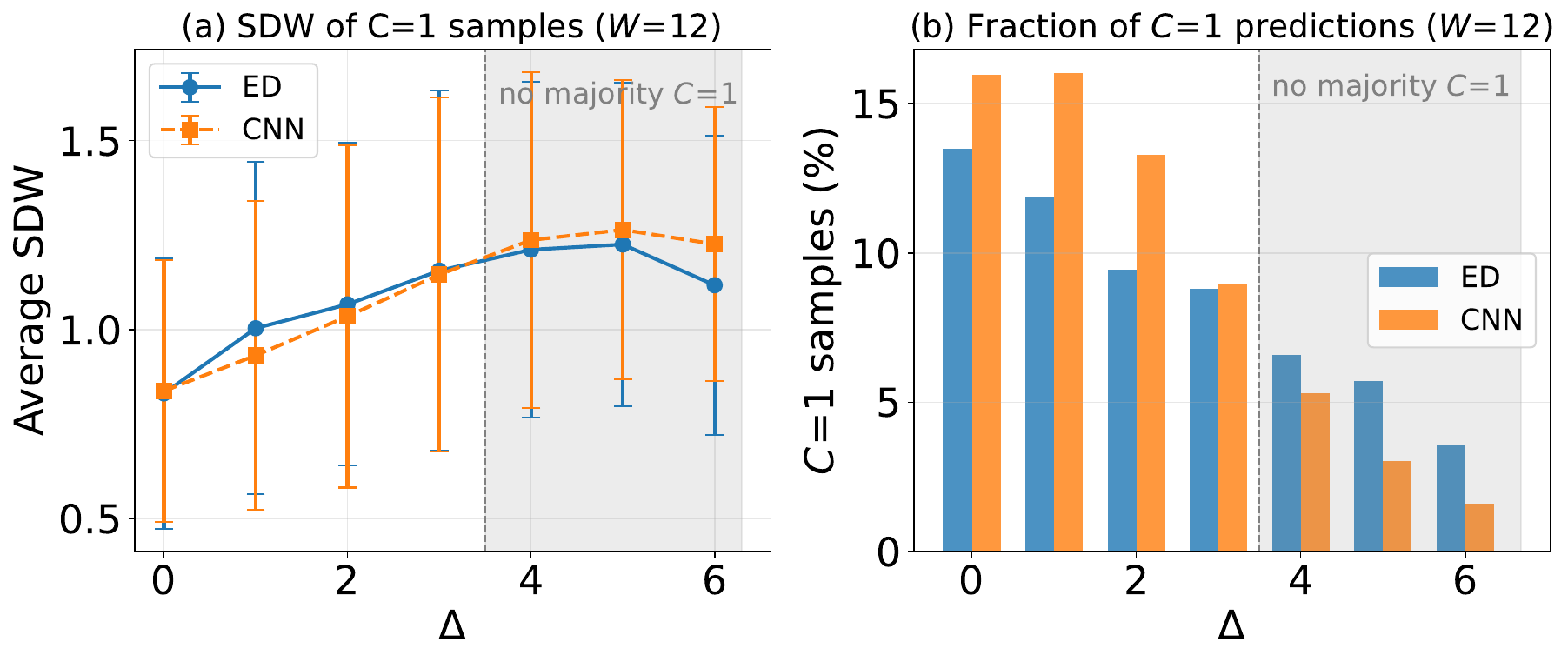}
    \caption{(a) Average SDW structure factor corresponding to the $C=1$ phase, as predicted by both ED and the CNN, as a function of the staggered mass. The error bars denote the standard deviation of the SDW factor over all $C=1$ samples, across $N_s=15$ disorder realization for ED and $N_s=100$ for the CNN. (b) Percentage of samples predicted to belong to the $C=1$ class, for both ED and the CNN, as a function of the staggered mass $\Delta$. In both panels, the grey region denotes values of $\Delta$ for which the $C=1$ phase does not appear as a majority phase.}
    \label{fig:fig6}
\end{figure}

To quantify the evolution of the $C=1$ phase as a function of the staggered mass, we compute the average SDW structure factor associated to the $C=1$ phase. Specifically, we average the SDW structure factor over all samples classified as $C=1$, both for ED and the CNN. The average SDW parameter is shown in Fig.~\ref{fig:fig6}(a) as a function of $\Delta$. We find that ED and CNN agree in the average value, as expected from their similar predicted topological regions. The grey region denotes values of $\Delta$ where there is no majority $C=1$ phase, although the phase can still appear for specific disorder realizations, leading to finite values of the SDW structure factor. This is shown in Fig.~\ref{fig:fig6}(b), where we plot the percentage of samples classified as belonging to the antiferromagnetic $C=1$ phase. We observe a smooth decline as a function of $\Delta$, and, as before, for $\Delta>3$ no majority $C=1$ phase is found, even though some seeds might display it. Importantly, the network tends to slightly overpredict the $C=1$ region. This is consistent with the reported $F_1$ score, obtained from a recall of $\sim 0.9$ and a precision of $\sim 0.9$, i.e. the network occasionally classifies $C=0$ or $C=2$ samples as $C=1$. This tendency can be seen also in Fig.~\ref{fig:fig5}, where the extent of the $C=1$ phase seems to be slightly larger in the CNN predictions than the ED results. Nevertheless, given the overall agreement between both methods, especially in light of the improved disorder statistics provided by the classifier, the CNN predictions give a useful estimate of the shape and position of the antiferrogmagnetic Chern phase. A more detailed ED calculation would be needed to determine the phase boundaries more precisely.

Analogous to the previous analysis, where $\Delta$ was varied at fixed $W$, one can now fix $\Delta$ and change $W$. Starting from the clean model at $\Delta=2$, which exhibits the $C=1$ phase as shown in Fig.~\ref{fig:fig1}, increasing $W$ leads to a progressive deformation and shrinkage of both topological regions (see Supplemental Material). We therefore conclude that the clean $C=1$ phase and the disorder-induced $C=1$ phase correspond to the same phase, since they are smoothly connected across the $(W,\Delta)$ plane and show the same defining properties, namely the same Chern number and SDW character. 

\section{Conclusions}

Using finite-size exact diagonalization, we have identified the emergence of a disorder-induced topological phase with $C=1$ and a finite staggered SDW structure factor, which we label an antiferromagnetic topological Anderson insulator. We further support the existence of this phase with a neural network trained on the ED data, allowing us to improve the disorder statistics and estimate more accurately the extent of the topological regions. We connect this antiferromagnetic disorder-induced phase at zero staggered mass with the well-established antiferromagnetic quantum anomalous Hall insulator of the clean model, which appears at finite staggered mass. Both phases appear to be smoothly connected as the staggered mass is varied, as seen from the evolution of the topological regions in the phase diagrams and from the behavior of the SDW structure factor.

The appearance of the antiferromagnetic Chern phase supports the previous conjecture \cite{hafez2025explicit} that an explicit charge order is necessary to induce this phase, here generated by the Anderson disorder term. In this sense, for the parameter values considered here, the staggered mass and the Anderson disorder appear to compete. A more detailed analysis of their interplay would be relevant to determine whether there exists a regime in which their effects can instead be constructive. The methods used here are limited by finite-size effects, finite disorder statistics, and the resolution of the parameter-space mesh. Hence, further studies with improved disorder sampling, finer meshes and larger system sizes would be valuable to verify and refine these findings.
Finally, since the antiferromagnetic $C=1$ phase is known to appear in several Chern insulator models, we conjecture that analogous antiferromagnetic topological Anderson $C=1$ phases may also appear upon the introduction of disorder.

\begin{acknowledgments} 
We thank Axel Fünfhaus for bringing relevant literature to our attention.
This work was supported by the Deutsche Forschungsgemeinschaft (DFG, German Research Foundation) through FOR5249 - 449872909 (Project P4) and TRR 288 —422213477 (Project A05).
\end{acknowledgments}

\bibliography{biblio}% Produces the bibliography via BibTeX.

\end{document}